\documentclass[lettersize,journal]{IEEEtran}

\usepackage{amsmath,amsfonts}
\usepackage{algorithmic}
\usepackage{algorithm}
\usepackage{array}
\usepackage{comment}
\usepackage{textcomp}
\usepackage{stfloats}
\usepackage{url}
\usepackage{verbatim}
\usepackage{cite}

\usepackage[acronyms,nonumberlist,nopostdot,nomain,nogroupskip,acronymlists={hidden}]{glossaries}
\newacronym{3gpp}{3GPP}{3rd Generation Partnership Project}
\newacronym{4g}{4G}{4th generation}
\newacronym{5g}{5G}{5th generation}
\newacronym{6g}{6G}{6th generation}
\newacronym{5gc}{5GC}{5G Core}
\newacronym{aau}{AAU}{Active Antenna Unit}
\newacronym{adc}{ADC}{Analog to Digital Converter}
\newacronym{aerpaw}{AERPAW}{Aerial Experimentation and Research Platform for Advanced Wireless}
\newacronym{ai}{AI}{Artificial Intelligence}
\newacronym{aimd}{AIMD}{Additive Increase Multiplicative Decrease}
\newacronym{am}{AM}{Acknowledged Mode}
\newacronym{amc}{AMC}{Adaptive Modulation and Coding}
\newacronym{amf}{AMF}{Access and Mobility Management Function}
\newacronym{aops}{AOPS}{Adaptive Order Prediction Scheduling}
\newacronym{api}{API}{Application Programming Interface}
\newacronym{apn}{APN}{Access Point Name}
\newacronym{ap}{AP}{Application Protocol}
\newacronym{aqm}{AQM}{Active Queue Management}
\newacronym{ausf}{AUSF}{Authentication Server Function}
\newacronym{avc}{AVC}{Advanced Video Coding}
\newacronym{awgn}{AGWN}{Additive White Gaussian Noise}
\newacronym{balia}{BALIA}{Balanced Link Adaptation Algorithm}
\newacronym{bbu}{BBU}{Base Band Unit}
\newacronym{bdp}{BDP}{Bandwidth-Delay Product}
\newacronym{ber}{BER}{Bit Error Rate}
\newacronym{bf}{BF}{Beamforming}
\newacronym{bler}{BLER}{Block Error Rate}
\newacronym{brr}{BRR}{Bayesian Ridge Regressor}
\newacronym{bs}{BS}{Base Station}
\newacronym{bsr}{BSR}{Buffer Status Report}
\newacronym{bss}{BSS}{Business Support System}
\newacronym{ca}{CA}{Carrier Aggregation}
\newacronym{caas}{CaaS}{Connectivity-as-a-Service}
\newacronym{cav}{CAV}{Connected and Autonoums Vehicle}
\newacronym{cb}{CB}{Code Block}
\newacronym{cc}{CC}{Congestion Control}
\newacronym{ccid}{CCID}{Congestion Control ID}
\newacronym{cco}{CC}{Carrier Component}
\newacronym{cd}{CD}{Continuous Delivery}
\newacronym{cdd}{CDD}{Cyclic Delay Diversity}
\newacronym{cdf}{CDF}{Cumulative Distribution Function}
\newacronym{cdn}{CDN}{Content Distribution Network}
\newacronym{cli}{CLI}{Command-line Interface}
\newacronym{cn}{CN}{Core Network}
\newacronym{codel}{CoDel}{Controlled Delay Management}
\newacronym{comac}{COMAC}{Converged Multi-Access and Core}
\newacronym{cord}{CORD}{Central Office Re-architected as a Datacenter}
\newacronym{cornet}{CORNET}{COgnitive Radio NETwork}
\newacronym{cosmos}{COSMOS}{Cloud Enhanced Open Software Defined Mobile Wireless Testbed for City-Scale Deployment}
\newacronym{cots}{COTS}{Commercial Off-the-Shelf}
\newacronym{cp}{CP}{Control Plane}
\newacronym{cpe}{CPE}{Customer Premises Equipment}
\newacronym{cyp}{CP}{Cyclic Prefix}
\newacronym{up}{UP}{User Plane}
\newacronym{cpu}{CPU}{Central Processing Unit}
\newacronym{cqi}{CQI}{Channel Quality Information}
\newacronym{cr}{CR}{Cognitive Radio}
\newacronym{cran}{CRAN}{Cloud \gls{ran}}
\newacronym{crs}{CRS}{Cell Reference Signal}
\newacronym{csi}{CSI}{Channel State Information}
\newacronym{csirs}{CSI-RS}{Channel State Information - Reference Signal}
\newacronym{cu}{CU}{Central Unit}
\newacronym{d2tcp}{D$^2$TCP}{Deadline-aware Data center TCP}
\newacronym{d3}{D$^3$}{Deadline-Driven Delivery}
\newacronym{dac}{DAC}{Digital to Analog Converter}
\newacronym{dag}{DAG}{Directed Acyclic Graph}
\newacronym{das}{DAS}{Distributed Antenna System}
\newacronym{dash}{DASH}{Dynamic Adaptive Streaming over HTTP}
\newacronym{dc}{DC}{Dual Connectivity}
\newacronym{dccp}{DCCP}{Datagram Congestion Control Protocol}
\newacronym{dce}{DCE}{Direct Code Execution}
\newacronym{dci}{DCI}{Downlink Control Information}
\newacronym{dctcp}{DCTCP}{Data Center TCP}
\newacronym{dl}{DL}{Downlink}
\newacronym{dmr}{DMR}{Deadline Miss Ratio}
\newacronym{dmrs}{DMRS}{DeModulation Reference Signal}
\newacronym{drlcc}{DRL-CC}{Deep Reinforcement Learning Congestion Control}
\newacronym{dsrc}{DSRC}
{dedicated short-range communications}
\newacronym{d2d}{D2D}{device-to-device}
\newacronym{drs}{DRS}{Discovery Reference Signal}
\newacronym{du}{DU}{Distributed Unit}
\newacronym{e2e}{E2E}{end-to-end}
\newacronym{earfcn}{EARFCN}{E-UTRA Absolute Radio Frequency Channel Number}
\newacronym{ecaas}{ECaaS}{Edge-Cloud-as-a-Service}
\newacronym{ecn}{ECN}{Explicit Congestion Notification}
\newacronym{edf}{EDF}{Earliest Deadline First}
\newacronym{embb}{eMBB}{Enhanced Mobile Broadband}
\newacronym{empower}{EMPOWER}{EMpowering transatlantic PlatfOrms for advanced WirEless Research}
\newacronym{enb}{eNB}{evolved Node Base}
\newacronym{endc}{EN-DC}{E-UTRAN-\gls{nr} \gls{dc}}
\newacronym{epc}{EPC}{Evolved Packet Core}
\newacronym{eps}{EPS}{Evolved Packet System}
\newacronym{es}{ES}{Edge Server}
\newacronym{etsi}{ETSI}{European Telecommunications Standards Institute}
\newacronym[firstplural=Estimated Times of Arrival (ETAs)]{eta}{ETA}{Estimated Time of Arrival}
\newacronym{eutran}{E-UTRAN}{Evolved Universal Terrestrial Access Network}
\newacronym{faas}{FaaS}{Function-as-a-Service}
\newacronym{fapi}{FAPI}{Functional Application Platform Interface}
\newacronym{fdd}{FDD}{Frequency Division Duplexing}
\newacronym{fdm}{FDM}{Frequency Division Multiplexing}
\newacronym{fdma}{FDMA}{Frequency Division Multiple Access}
\newacronym{fed4fire}{FED4FIRE+}{Federation 4 Future Internet Research and Experimentation Plus}
\newacronym{fir}{FIR}{Finite Impulse Response}
\newacronym{fit}{FIT}{Future \acrlong{iot}}
\newacronym{fpga}{FPGA}{Field Programmable Gate Array}
\newacronym{fr2}{FR2}{Frequency Range 2}
\newacronym{fr1}{FR1}{Frequency Range 1}
\newacronym{fs}{FS}{Fast Switching}
\newacronym{fscc}{FSCC}{Flow Sharing Congestion Control}
\newacronym{ftp}{FTP}{File Transfer Protocol}
\newacronym{fw}{FW}{Flow Window}
\newacronym{ge}{GE}{Gaussian Elimination}
\newacronym{gnb}{gNB}{Next Generation Node Base}
\newacronym{gop}{GOP}{Group of Pictures}
\newacronym{gpr}{GPR}{Gaussian Process Regressor}
\newacronym{gpu}{GPU}{Graphics Processing Unit}
\newacronym{gtp}{GTP}{GPRS Tunneling Protocol}
\newacronym{gtpc}{GTP-C}{GPRS Tunnelling Protocol Control Plane}
\newacronym{gtpu}{GTP-U}{GPRS Tunnelling Protocol User Plane}
\newacronym{gtpv2c}{GTPv2-C}{\gls{gtp} v2 - Control}
\newacronym{gw}{GW}{Gateway}
\newacronym{harq}{HARQ}{Hybrid Automatic Repeat reQuest}
\newacronym{hetnet}{HetNet}{Heterogeneous Network}
\newacronym{hh}{HH}{Hard Handover}
\newacronym{hol}{HOL}{Head-of-Line}
\newacronym{hqf}{HQF}{Highest-quality-first}
\newacronym{hss}{HSS}{Home Subscription Server}
\newacronym{http}{HTTP}{HyperText Transfer Protocol}
\newacronym{ia}{IA}{Initial Access}
\newacronym{iab}{IAB}{Integrated Access and Backhaul}
\newacronym{ic}{IC}{Incident Command}
\newacronym{ietf}{IETF}{Internet Engineering Task Force}
\newacronym{ieee}{IEEE}{Institute of Electrical and Electronics Engineers}
\newacronym{imsi}{IMSI}{International Mobile Subscriber Identity}
\newacronym{iot}{IoT}{Internet of Things}
\newacronym{ip}{IP}{Internet Protocol}
\newacronym{isac}{ISAC}{Integrated Sensing and Communication}
\newacronym{itu}{ITU}{International Telecommunication Union}
\newacronym{kpi}{KPI}{Key Performance Indicator}
\newacronym{kpm}{KPM}{Key Performance Measurement}
\newacronym{kvm}{KVM}{Kernel-based Virtual Machine}
\newacronym{los}{LoS}{Line of Sight}
\newacronym{lsm}{LSM}{Link-to-System Mapping}
\newacronym{lstm}{LSTM}{Long Short Term Memory}
\newacronym{lte}{LTE}{Long Term Evolution}
\newacronym{lxc}{LXC}{Linux Container}
\newacronym{m2m}{M2M}{Machine to Machine}
\newacronym{mac}{MAC}{Medium Access Control}
\newacronym{manet}{MANET}{Mobile Ad Hoc Network}
\newacronym{mano}{MANO}{Management and Orchestration}
\newacronym{mc}{MC}{Multi-Connectivity}
\newacronym{mcc}{MCC}{Mobile Cloud Computing}
\newacronym{mchem}{MCHEM}{Massive Channel Emulator}
\newacronym{mcs}{MCS}{Modulation and Coding Scheme}
\newacronym{mec2}{MEC}{Multi-access Edge Computing}
\newacronym{mec}{MEC}{Mobile Edge Computing}
\newacronym{mfc}{MFC}{Mobile Fog Computing}
\newacronym{mgen}{MGEN}{Multi-Generator}
\newacronym{mi}{MI}{Mutual Information}
\newacronym{mib}{MIB}{Master Information Block}
\newacronym{miesm}{MIESM}{Mutual Information Based Effective SINR}
\newacronym{mimo}{MIMO}{Multiple Input, Multiple Output}
\newacronym{ml}{ML}{Machine Learning}
\newacronym{mlr}{MLR}{Maximum-local-rate}
\newacronym[plural=\gls{mme}s,firstplural=Mobility Management Entities (MMEs)]{mme}{MME}{Mobility Management Entity}
\newacronym{mmtc}{mMTC}{Massive Machine-Type Communications}
\newacronym{mmwave}{mmWave}{millimeter wave}
\newacronym{mno}{MNO}{Mobile Network Operators}
\newacronym{mpdccp}{MP-DCCP}{Multipath Datagram Congestion Control Protocol}
\newacronym{mptcp}{MPTCP}{Multipath TCP}
\newacronym{mr}{MR}{Maximum Rate}
\newacronym{mrn}{MRN}{Mobile Radio Network}
\newacronym{mrdc}{MR-DC}{Multi \gls{rat} \gls{dc}}
\newacronym{mse}{MSE}{Mean Square Error}
\newacronym{mss}{MSS}{Maximum Segment Size}
\newacronym{mt}{MT}{Mobile Termination}
\newacronym{mtd}{MTD}{Machine-Type Device}
\newacronym{mtu}{MTU}{Maximum Transmission Unit}
\newacronym{mumimo}{MU-MIMO}{Multi-user \gls{mimo}}
\newacronym{mvno}{MVNO}{Mobile Virtual Network Operator}
\newacronym{nalu}{NALU}{Network Abstraction Layer Unit}
\newacronym{nas}{NAS}{Network Attached Storage}
\newacronym{nat}{NAT}{Network Address Translation}
\newacronym{nbiot}{NB-IoT}{Narrow Band IoT}
\newacronym{nfv}{NFV}{Network Function Virtualization}
\newacronym{nfvi}{NFVI}{Network Function Virtualization Infrastructure}
\newacronym{ni}{NI}{Network Interfaces}
\newacronym{nic}{NIC}{Network Interface Card}
\newacronym{now}{NOW}{Non Overlapping Window}
\newacronym{nsm}{NSM}{Network Service Mesh}
\newacronym{nr}{NR}{New Radio}
\newacronym{nrf}{NRF}{Network Repository Function}
\newacronym{nr-u}{NR-U}{New Radio Unlicensed}
\newacronym{nsa}{NSA}{Non Stand Alone}
\newacronym{nse}{NSE}{Network Slicing Engine}
\newacronym{nssf}{NSSF}{Network Slice Selection Function}
\newacronym{oai}{OAI}{OpenAirInterface}
\newacronym{oaicn}{OAI-CN}{\gls{oai} \acrlong{cn}}
\newacronym{oairan}{OAI-RAN}{\acrlong{oai} \acrlong{ran}}
\newacronym{oam}{OAM}{Operations, Administration and Maintenance}
\newacronym{ofdm}{OFDM}{Orthogonal Frequency Division Multiplexing}
\newacronym{olia}{OLIA}{Opportunistic Linked Increase Algorithm}
\newacronym{omec}{OMEC}{Open Mobile Evolved Core}
\newacronym{onap}{ONAP}{Open Network Automation Platform}
\newacronym{onf}{ONF}{Open Networking Foundation}
\newacronym{onos}{ONOS}{Open Networking Operating System}
\newacronym{oom}{OOM}{\gls{onap} Operations Manager}
\newacronym{opnfv}{OPNFV}{Open Platform for \gls{nfv}}
\newacronym{Oran}{O-RAN}{Open \gls{ran}}
\newacronym{orbit}{ORBIT}{Open-Access Research Testbed for Next-Generation Wireless Networks}
\newacronym{os}{OS}{Operating System}
\newacronym{oss}{OSS}{Operations Support System}
\newacronym{pa}{PA}{Position-aware}
\newacronym{pase}{PASE}{Prioritization, Arbitration, and Self-adjusting Endpoints}
\newacronym{pawr}{PAWR}{Platforms for Advanced Wireless Research}
\newacronym{pbch}{PBCH}{Physical Broadcast Channel}
\newacronym{pcef}{PCEF}{Policy and Charging Enforcement Function}
\newacronym{pcfich}{PCFICH}{Physical Control Format Indicator Channel}
\newacronym{pcrf}{PCRF}{Policy and Charging Rules Function}
\newacronym{pdcch}{PDCCH}{Physical Downlink Control Channel}
\newacronym{pdcp}{PDCP}{Packet Data Convergence Protocol}
\newacronym{pdsch}{PDSCH}{Physical Downlink Shared Channel}
\newacronym{pdu}{PDU}{Packet Data Unit}
\newacronym{pf}{PF}{Proportional Fair}
\newacronym{pgw}{PGW}{Packet Gateway}
\newacronym{phich}{PHICH}{Physical Hybrid ARQ Indicator Channel}
\newacronym{phy}{PHY}{Physical}
\newacronym{pmch}{PMCH}{Physical Multicast Channel}
\newacronym{pmi}{PMI}{Precoding Matrix Indicators}
\newacronym{powder}{POWDER}{Platform for Open Wireless Data-driven Experimental Research}
\newacronym{ppo}{PPO}{Proximal Policy Optimization}
\newacronym{ppp}{PPP}{Poisson Point Process}
\newacronym{prach}{PRACH}{Physical Random Access Channel}
\newacronym{prb}{PRB}{Physical Resource Block}
\newacronym{psnr}{PSNR}{Peak Signal to Noise Ratio}
\newacronym{pss}{PSS}{Primary Synchronization Signal}
\newacronym{pucch}{PUCCH}{Physical Uplink Control Channel}
\newacronym{pusch}{PUSCH}{Physical Uplink Shared Channel}
\newacronym{qam}{QAM}{Quadrature Amplitude Modulation}
\newacronym{qci}{QCI}{\gls{qos} Class Identifier}
\newacronym{qoe}{QoE}{Quality of Experience}
\newacronym{qos}{QoS}{Quality of Service}
\newacronym{quic}{QUIC}{Quick UDP Internet Connections}
\newacronym{ra}{RA}{Resouces Allocation}
\newacronym{rach}{RACH}{Random Access Channel}
\newacronym{ran}{RAN}{Radio Access Network}
\newacronym[firstplural=Radio Access Technologies (RATs)]{rat}{RAT}{Radio Access Technology}
\newacronym{rbg}{RBG}{Resource Block Group}
\newacronym{rcn}{RCN}{Research Coordination Network}
\newacronym{rc}{RC}{RAN Control}
\newacronym{rec}{REC}{Radio Edge Cloud}
\newacronym{red}{RED}{Random Early Detection}
\newacronym{renew}{RENEW}{Reconfigurable Eco-system for Next-generation End-to-end Wireless}
\newacronym{rf}{RF}{Radio Frequency}
\newacronym{rfc}{RFC}{Request for Comments}
\newacronym{rfr}{RFR}{Random Forest Regressor}
\newacronym{ric}{RIC}{\gls{ran} Intelligent Controller}
\newacronym{rim}{RIM}{Remote Interference Management}
\newacronym{rlc}{RLC}{Radio Link Control}
\newacronym{rlf}{RLF}{Radio Link Failure}
\newacronym{rlnc}{RLNC}{Random Linear Network Coding}
\newacronym{rmr}{RMR}{RIC Message Router}
\newacronym{rmse}{RMSE}{Root Mean Squared Error}
\newacronym{rnis}{RNIS}{Radio Network Information Service}
\newacronym{rr}{RR}{Round Robin}
\newacronym{rrc}{RRC}{Radio Resource Control}
\newacronym{rrm}{RRM}{Radio Resource Management}
\newacronym{rru}{RRU}{Remote Radio Unit}
\newacronym{rs}{RS}{Remote Server}
\newacronym{rsrp}{RSRP}{Reference Signal Received Power}
\newacronym{rsrq}{RSRQ}{Reference Signal Received Quality}
\newacronym{rss}{RSS}{Received Signal Strength}
\newacronym{rssi}{RSSI}{Received Signal Strength Indicator}
\newacronym{rtt}{RTT}{Round Trip Time}
\newacronym{ru}{RU}{Radio Unit}
\newacronym{rus}{RSU}{Road Side Unit}
\newacronym{rw}{RW}{Receive Window}
\newacronym{rx}{RX}{Receiver}
\newacronym{s1ap}{S1AP}{S1 Application Protocol}
\newacronym{sa}{SA}{standalone}
\newacronym{sack}{SACK}{Selective Acknowledgment}
\newacronym{sap}{SAP}{Service Access Point}
\newacronym{sc2}{SC2}{Spectrum Collaboration Challenge}
\newacronym{scef}{SCEF}{Service Capability Exposure Function}
\newacronym{sch}{SCH}{Secondary Cell Handover}
\newacronym{scs}{SCS}{Sub-Carrier Spacing}
\newacronym{scoot}{SCOOT}{Split Cycle Offset Optimization Technique}
\newacronym{sctp}{SCTP}{Stream Control Transmission Protocol}
\newacronym{sdap}{SDAP}{Service Data Adaptation Protocol}
\newacronym{sdk}{SDK}{Software Development Kit}
\newacronym{sdm}{SDM}{Space Division Multiplexing}
\newacronym{sdma}{SDMA}{Spatial Division Multiple Access}
\newacronym{sdn}{SDN}{Software-defined Networking}
\newacronym{sdr}{SDR}{Software-defined Radio}
\newacronym{seba}{SEBA}{SDN-Enabled Broadband Access}
\newacronym{sgsn}{SGSN}{Serving GPRS Support Node}
\newacronym{sgw}{SGW}{Service Gateway}
\newacronym{si}{SI}{Study Item}
\newacronym{sib}{SIB}{Secondary Information Block}
\newacronym{sinr}{SINR}{Signal to Interference plus Noise Ratio}
\newacronym{sip}{SIP}{Session Initiation Protocol}
\newacronym{siso}{SISO}{Single Input, Single Output}
\newacronym{sla}{SLA}{Service Level Agreement}
\newacronym{sm}{SM}{Service Model}
\newacronym{smo}{SMO}{Service Management and Orchestration}
\newacronym{smsgmsc}{SMS-GMSC}{\gls{sms}-Gateway}
\newacronym{snr}{SNR}{Signal-to-Noise-Ratio}
\newacronym{son}{SON}{Self-Organizing Network}
\newacronym{sptcp}{SPTCP}{Single Path TCP}
\newacronym{srb}{SRB}{Service Radio Bearer}
\newacronym{srn}{SRN}{Standard Radio Node}
\newacronym{srs}{SRS}{Sounding Reference Signal}
\newacronym{ss}{SS}{Synchronization Signal}
\newacronym{ssb}{SSB}{Synchronization Signal Block}
\newacronym{sss}{SSS}{Secondary Synchronization Signal}
\newacronym{st}{ST}{Spanning Tree}
\newacronym{svc}{SVC}{Scalable Video Coding}
\newacronym{tb}{TB}{Transport Block}
\newacronym{tcp}{TCP}{Transmission Control Protocol}
\newacronym{tdd}{TDD}{Time Division Duplexing}
\newacronym{tdm}{TDM}{Time Division Multiplexing}
\newacronym{tdma}{TDMA}{Time Division Multiple Access}
\newacronym{tfl}{TfL}{Transport for London}
\newacronym{tfrc}{TFRC}{TCP-Friendly Rate Control}
\newacronym{tft}{TFT}{Traffic Flow Template}
\newacronym{tgen}{TGEN}{Traffic Generator}
\newacronym{tip}{TIP}{Telecom Infra Project}
\newacronym{tm}{TM}{Transparent Mode}
\newacronym{to}{TO}{Telco Operator}
\newacronym{tr}{TR}{Technical Report}
\newacronym{trp}{TRP}{Transmitter Receiver Pair}
\newacronym{ts}{TS}{Technical Specification}
\newacronym{tti}{TTI}{Transmission Time Interval}
\newacronym{ttt}{TTT}{Time-to-Trigger}
\newacronym{tue}{TUE}{Test UE}
\newacronym{tx}{TX}{Transmitter}
\newacronym{u6g}{U6G}{Upper 6GHz}
\newacronym{uas}{UAS}{Unmanned Aerial System}
\newacronym{uav}{UAV}{Unmanned Aerial Vehicle}
\newacronym{udm}{UDM}{Unified Data Management}
\newacronym{udp}{UDP}{User Datagram Protocol}
\newacronym{udr}{UDR}{Unified Data Repository}
\newacronym{ue}{UE}{User Equipment}
\newacronym{uhd}{UHD}{\gls{usrp} Hardware Driver}
\newacronym{ul}{UL}{Uplink}
\newacronym{um}{UM}{Unacknowledged Mode}
\newacronym{uml}{UML}{Unified Modeling Language}
\newacronym{upa}{UPA}{Uniform Planar Array}
\newacronym{upf}{UPF}{User Plane Function}
\newacronym{urllc}{URLLC}{Ultra Reliable and Low Latency Communications}
\newacronym{usa}{U.S.}{United States}
\newacronym{usim}{USIM}{Universal Subscriber Identity Module}
\newacronym{usrp}{USRP}{Universal Software Radio Peripheral}
\newacronym{utc}{UTC}{Urban Traffic Control}
\newacronym{vim}{VIM}{Virtualization Infrastructure Manager}
\newacronym{vm}{VM}{Virtual Machine}
\newacronym{vnf}{VNF}{Virtual Network Function}
\newacronym{volte}{VoLTE}{Voice over \gls{lte}}
\newacronym{voltha}{VOLTHA}{Virtual OLT HArdware Abstraction}
\newacronym{vr}{VR}{Virtual Reality}
\newacronym{vran}{vRAN}{Virtualized \gls{ran}}
\newacronym{vss}{VSS}{Video Streaming Server}
\newacronym{v2x}{V2X}{vehicle-to-everything}
\newacronym{v2i}{V2I}{vehicle-to-infrastructure}
\newacronym{v2v}{V2V}{vehicle-to-vehicle}
\newacronym{v2n}{V2N}{vehicle-to-network}
\newacronym{wbf}{WBF}{Wired Bias Function}
\newacronym{wf}{WF}{Waterfilling}
\newacronym{wg}{WG}{Working Group}
\newacronym{wlan}{WLAN}{Wireless Local Area Network}
\newacronym{wrc}{WRC}{World Radiocommunication Conference}
\newacronym{osm}{OSM}{Open Source \gls{nfv} Management and Orchestration}
\newacronym{pnf}{PNF}{Physical Network Function}
\newacronym{drl}{DRL}{Deep Reinforcement Learning}
\newacronym{mtc}{MTC}{Machine-type Communications}
\newacronym{osc}{OSC}{O-RAN Software Community}
\newacronym{mns}{MnS}{Management Services}
\newacronym{ves}{VES}{\gls{vnf} Event Stream}
\newacronym{ei}{EI}{Enrichment Information}
\newacronym{fh}{FH}{Fronthaul}
\newacronym{fft}{FFT}{Fast Fourier Transform}
\newacronym{laa}{LAA}{Licensed-Assisted Access}
\newacronym{plfs}{PLFS}{Physical Layer Frequency Signals}
\newacronym{ptp}{PTP}{Precision Time Protocol}
\newacronym{lidar}{LiDAR}{Light Detection And Ranging}
\newacronym{dem}{DEM}{Digital Elevation Model}
\newacronym{dtm}{DEM}{Digital Terrain Model}
\newacronym{dsm}{DEM}{Digital Surface Models}
\newacronym{ota}{OTA}{Over-The-Air}
\newacronym{ns}{NS}{Network Slicing}
\newacronym{ne}{NE}{Nash Equilibrium}
\newacronym{hf}{HF}{High Frequency}
\newacronym{noma}{NOMA}{Non-Orthogonal Multiple Access}
\newacronym{sre}{SRE}{Smart Radio Environment}
\newacronym{ris}{RIS}{Reconfigurable Intelligent Surface}
\newacronym{inp}{InP}{Infrastructure Provider}
\newacronym{smf}{SMF}{Slicing Magangement Framework}
\newacronym{nsn}{NSN}{Network Slicing Negotiation}
\newacronym{sms}{SMS}{Slicing MAC Scheduler}
\newacronym{brd}{BRD}{Best Response Dynamics}
\newacronym{dssbr}{DSSBR}{Double Step Smoothed Best Response}
\newacronym{poa}{PoA}{Price of Anarchy}
\newacronym{pos}{PoS}{Price of Stability}
\newacronym{milp}{MILP}{Mixed Integer-Linear Program}
\newacronym{pod}{PoD}{Price of DSSBR}
\newacronym{roc}{ROC}{Radio Overload Control}
\newacronym{ciot}{cIoT}{critical Internet of Things}
\newacronym{embbpr}{eMBB Pr.}{enhanced Mobile BroadBand Premium}
\newacronym{sps}{SPS}{Semi-persistent Scheduling}
\newacronym{cg}{CG}{Configured Grant}
\newacronym{embbbs}{eMBB Bs.}{enhanced Mobile BroadBand Basic}
\newacronym{en}{EN}{Edge Node}
\newacronym{ec}{EC}{Edge Computing}
\newacronym{sp}{SP}{Service Provider}
\newacronym{me}{ME}{Market Equilibrium}
\newacronym{so}{SO}{Social Optimum}
\newacronym{wso}{WSO}{Weighted Social Optimum}
\newacronym{ps}{PS}{Proportional Sharing}
\newacronym{eg}{EG}{Eisenberg-Gale program}
\newacronym{pe}{PE}{Pareto Efficiency}
\newacronym{nsw}{NSW}{Nash Social Welfare}
\newacronym{ef}{EF}{Envy-Freeness}
\newacronym{sub6}{sub6GHz}{Below 6GHz}
\newacronym{ncr}{NCR}{Network-Controlled Repeater}
\newacronym{nlos}{NLoS}{Non-Line of Sight}
\newacronym{ntp}{NTP}{Network Time Protocol}
\newacronym{src}{SRC}{Smart Radio Connection}
\newacronym{srd}{SRD}{Smart Radio Device}
\newacronym{cs}{CS}{Candidate Site}
\newacronym{tp}{TP}{Test Point}
\newacronym{fov}{FoV}{Field of View}
\newacronym{nrric}{near-RT RIC}{Near Real-time {RAN} Intelligent Controller}
\newacronym{e2ap}{E2AP}{E2 Application Protocol}
\newacronym{e2sm}{E2SM}{E2 Service Model}
\newacronym{nrtric}{non-RT RIC}{Non-Real-Time {RIC}}
\newacronym{itti}{ITTI}{Inter-task Interface}
\newacronym{bap}{BAP}{Backhaul Adaptation Protocol}
\newacronym{iabest}{IABEST}{Integrated Access and Backhaul Experimental large-Scale Tetbed}
\newacronym{teid}{TEID}{Tunnel Endpoint Identifier}
\newacronym{dlsch}{DL-SCH}{Downlink Shared Channel }
\newacronym{ulsch}{UL-SCH}{Uplink Shared Channel }
\newacronym{rsu}{RSU}{Road Side Unit}
\newacronym{its}{ITS}{Intelligent Transportation Systems}
\newacronym{vanet}{VANET}{Vehicular Ad-hoc Network}
\newacronym{dt}{DT}{Digital Twin}
\newacronym{ecc}{ECC}{Edge Computing Cluster}
\newacronym{o2i}{O2I}{Outdoor-to-indoor}
\newacronym{fwa}{FWA}{Fixed Wireless Access}
\newacronym{afc}{AFC}{Automated Frequency Coordinator}
\newacronym{o2o}{O2O}{Outdoor-to-outdoor}
\newacronym{hpbw}{HPBW}{Half-Power Beamwidth}
\newacronym{rb}{RB}{Resource Block}
\newacronym{cca}{CCA}{Congestion Control Algorithm}
\newacronym{wab}{WAB}{Wireless Access and Backhaul}
\newacronym{mwab}{MWAB}{Mobile Wireless Access and Backhaul}
\newacronym{ntn}{NTN}{Non-Terrestrial Networks}
\newacronym{plmn}{PLMN}{Public Land Mobile Network}
\newacronym{mbsr}{MBSR}{Mobile Base Station Relay}
\newacronym{vpn}{VPN}{Virtual Private Network}
\newacronym{sdwan}{SD-WAN}{Software-Defined Wide Area Network}
\newacronym{cbrs}{CBRS}{Citizens Broadband Radio Service}
\usepackage{subcaption}
\usepackage[]{graphicx}
\usepackage{svg}
\usepackage{overpic}
\usepackage[section]{placeins}
\newsavebox{\measurebox}
\usepackage{multirow}
\usepackage{hhline}
\usepackage{needspace}

\usepackage{pgfplots}
\usepackage{pgfplotstable}

\usepackage{stfloats}
\usepackage[table]{xcolor}

\usepackage{enumitem}

\usepackage{zref-xr}
\zexternaldocument{main}

\newcommand{\rev}[1]{{\color{black}#1}}

\begin{document}
%
\title{Toward Mobile and Converged Backhaul:\\ The Promise of Wireless Access and Backhaul}
%
%
%

\author{
Chiara~Rubaltelli,~\IEEEmembership{Member,~IEEE,}
Marcello~Morini,~\IEEEmembership{Member,~IEEE,}
Eugenio~Moro,~\IEEEmembership{Member,~IEEE,}
Ilario~Filippini,~\IEEEmembership{Senior Member,~IEEE,} and
Antonio~Capone,~\IEEEmembership{Fellow,~IEEE}
\vspace{-1cm}
\thanks{The authors are with Dipartimento di Elettronica, Informazione e Bioingegneria of Politecnico di Milano, Milan, Italy e-mail: \textit{name}.\textit{surname}@polimi.it}
}

%
%

\markboth{}%
{Shell \MakeLowercase{\textit{et al.}}: Bare Demo of IEEEtran.cls for IEEE Journals}
%



\maketitle

\begin{abstract}
\gls{wab} is emerging as a key enabler for flexible and cost-efficient 5G deployments, offering a modular architecture that decouples access and backhaul while supporting multi-technology and mobile backhaul links. This article introduces the \gls{wab} framework standardized in 3GPP Release~19, outlining its architecture and operational principles. A practical implementation built with commercial hardware and open-source software demonstrates the feasibility and efficiency of \gls{wab} systems. We further explore four representative application scenarios -- ranging from on-demand coverage to mobile \gls{sdwan} connectivity -- and discuss the technical challenges that must be addressed for large-scale adoption. These insights highlight \gls{wab} as a promising foundation for 5G-Advanced and a stepping stone toward future 6G networks.

\label{sec:abstract}   
\end{abstract}


\glsresetall

\vspace{-0.5cm}
\section{Introduction}
\label{sec:introduction}
\rev{Providing ubiquitous mobile coverage in a cost efficient and scalable manner remains a fundamental challenge for cellular networks. To address this, \gls{3gpp} has long explored relay-based solutions as a promising approach~\cite{relay_lte}. Early implementations emerged as home-based femtocells and later evolved into small cells within \gls{hetnet} architectures. Femtocells~\cite{femto}, introduced in the early 2000s, offered a simple and cost effective way to improve indoor coverage and offload traffic from macrocells. However, their reliance on conventional backhaul solutions, typically wired or satellite based, inherently limited their flexibility and hindered large scale adoption.

This limitation motivated the evolution toward wireless backhaul, culminating in the \gls{iab} framework introduced in Release~15~\cite{miab}. \Gls{iab} extends coverage through multi-hop tree topologies in which relay nodes can serve both downstream nodes and \glspl{ue}. While this represents a significant step forward, it preserves a tight coupling between access and backhaul. This coupling increases the complexity of resource management, interference coordination, and control plane operation, ultimately constraining flexibility and deployment scalability, and preventing widespread adoption.

The introduction of \gls{wab} in Release~19~\cite{wabtechreport38799, mwabtechreport23700} marks a decisive shift in this design space. Unlike femtocells and \gls{iab}, \gls{wab} decouples access and backhaul and adopts a modular architecture in which connectivity is provided through wireless backhaul links that can leverage both \gls{3gpp} and non-\gls{3gpp} technologies~\cite{5GAdvanced}. This enables seamless integration of heterogeneous networks, including mmWave and \gls{ntn} systems, whose direct support in commercial \glspl{ue} remains limited, by relocating their role to the backhaul segment. Moreover, \gls{wab} is inherently designed to support mobility at the backhaul level, enabling \glspl{mrn} where entire network segments can move while maintaining connectivity, thereby defining a new class of moving networks~\cite{moving_net}.

In this perspective, \gls{wab} is not merely an incremental evolution of relay-based solutions, but a fundamental rethinking of how access and backhaul are composed and operated. This paradigm shift unlocks unprecedented flexibility, while simultaneously introducing new technical challenges, including radio access reconfiguration under unreliable backhaul conditions, joint access and backhaul mobility management, and intelligent network and interference management. These challenges open unexplored research directions and call for new design principles.

Given that \gls{wab} is approaching standardization and that, to the best of our knowledge, it has not yet been systematically analyzed in the literature, this article provides a timely and comprehensive first overview of the technology. In particular, beyond presenting the architecture and a practical implementation, we place a strong emphasis on the analysis of the key technical challenges introduced by \gls{wab} and on the identification of the research directions they enable. We complement this discussion with experimental results obtained from a real world testbed, including mmWave-based backhaul, demonstrating the feasibility and performance of \gls{wab} deployments based on \gls{cots} hardware and open-source software. Specifically, the article aims to:
\begin{itemize}[leftmargin=14pt]
	\item outline the key technical features and benefits of \gls{wab};
	\item present and validate a practical implementation of \gls{wab};
	\item discuss the fundamental technical challenges and emerging research directions;
	\item describe representative application scenarios enabled by \gls{wab}.
\end{itemize}}

\vspace{-0.25cm}
\section{Wireless Access and Backhaul in \gls{3gpp}}
\label{sec:wab_3gpp}
\begin{figure*}[ht]
    \centering
    \includegraphics[width=1.6\columnwidth]{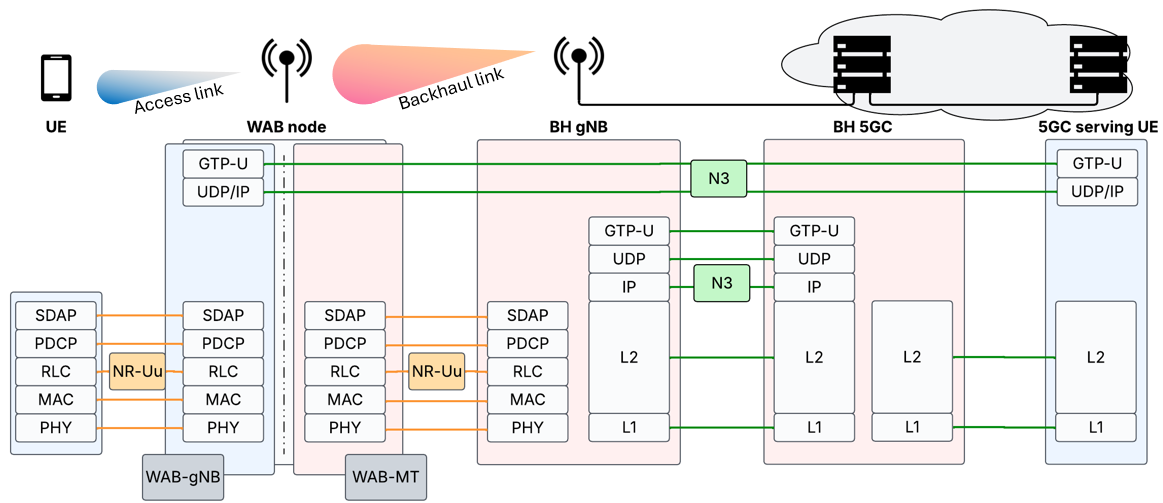}
    \caption{\gls{wab} architecture and protocol stack on the user plane. In light-blue the elements forming the access segment, while in red the elements composing the backhaul segment.\vspace{-0.5cm}}
    \label{fig:stack}
\end{figure*}

In late 2023, the \gls{3gpp} launched two \glspl{si} related to \gls{wab}, which were later adopted into Release~19 specifications as a topological enhancement for 5G-Advanced (5G-A). Within the Technical Specification Group (TSG) \gls{ran}, the \gls{wab} technology was introduced and technically assessed~\cite{wabtechreport38799}, while the Service and System Aspects (SA) group addressed the associated management framework, termed \gls{mwab}~\cite{mwabtechreport23700}. \rev{3GPP standardization efforts mainly target the definition of the architecture and configuration procedures, partially addressing performance optimization in the case of 3GPP backhaul.}

A primary motivation for this work is the increasing momentum of \gls{ntn} in the 6G standardization process. \Gls{wab} nodes are expected to play a central role in enabling non-terrestrial backhaul by providing 5G access in areas lacking terrestrial network (TN) coverage~\cite{ntn_architectures, RanPlenaryWAB}. This includes support for mobile backhaul links and seamless \gls{ntn}--TN handovers.
In addition, \gls{wab} has been envisioned as a facilitator for roaming \glspl{ue}, enabling them to connect to foreign networks by attaching to hybrid \gls{wab} nodes that provide home-network radio access. This approach removes the need for \gls{ran} sharing, as well as formal roaming agreements between operators.

%



\vspace{-0.25cm}
\subsection*{Architecture}
The \gls{wab} architecture comprises two communication systems -- the access segment and the backhaul segment -- capable of operating in either in-band or out-of-band configurations. End-to-end connectivity is achieved by tunneling the fundamental 3GPP interfaces that interconnect the functional elements of the access segment across the links of the backhaul segment. A schematic representation of the basic architecture, along with the corresponding user-plane protocol stacks, is shown in Figure~\ref{fig:stack}.

At the core of the architecture lies the \gls{wab} Node, which integrates two main components: a WAB \gls{gnb} (\textit{WAB-gNB}) and a WAB Mobile Termination (\textit{WAB-MT}). The \textit{WAB-gNB} is a 5G-compliant \gls{gnb} that provides radio access to end \glspl{ue}. The \textit{WAB-MT}, in the basic configuration, establishes a wireless link with a \textit{BH-gNB}, and this link is then used by the \textit{WAB-gNB} as a backhaul connection to forward \gls{ue} traffic -- tunneled by the \textit{WAB-gNB} -- toward the access network’s core, referred to as the \textit{5GC-serving-UE}. It is worth noting that, according to current \gls{3gpp} specifications, WAB Nodes are not permitted to obtain backhaul connectivity from other WAB Nodes, thereby preventing multi-hop topologies among them. \rev{This simplifies backhaul handovers and reduces procedural complexity, while avoiding issues related to insufficient resources to support the required backhaul \gls{pdu} sessions.}

Within the backhaul segment, the \textit{BH-gNB} connects to its core, \textit{BH-5GC}, via conventional transport technologies. The \textit{WAB-MT}, \textit{BH-gNB}, and \textit{BH-5GC} jointly establish a \gls{pdu} session responsible for maintaining connectivity between access-network interfaces, providing a transparent, tunneled backhaul transport service. Finally, the \textit{BH-5GC} connects to the \textit{5GC-serving-UE}, completing the end-to-end architecture.

On the control plane, the wireless backhaul is established through a backhaul-segment \gls{pdu} session that carries the N2 interface -- including the Next Generation Application Protocol (NGAP), \gls{sctp}, and \gls{ip} protocol layers -- connecting the \textit{WAB-gNB} to the \textit{5GC-serving-UE} \gls{amf}. On the user plane, a separate backhaul \gls{pdu} session supports the N3 interface -- including \gls{gtpu}, \gls{udp}, and \gls{ip} layers -- linking the \textit{WAB-gNB} with the \textit{5GC-serving-UE} \gls{upf}.

While, in the basic configuration, the access and backhaul segments operate as two standard-compliant, logically independent 5G networks, the \gls{wab} architecture also allows the backhaul segment to be implemented using non-\gls{3gpp} systems, unconventional networks, or even wired technologies. In all cases, the end \gls{ue} remains agnostic to the specific implementation of the backhaul.

\vspace{-0.35cm}
\subsection*{Basic Functioning and Procedures}

The integration of a \gls{wab} node into an existing network begins with the setup of the \textit{WAB-MT}. This component connects to the \textit{BH-gNB} as a standard 5G \gls{ue}, following conventional access procedures. Once authorized by the \textit{BH-5GC}, the \textit{WAB-MT} can establish one or more backhaul \gls{pdu} sessions with the \textit{BH-5GC}'s \gls{upf}. Multiple backhaul \gls{pdu} sessions may be created to transport different logical interfaces or to enable differentiated traffic handling according to specific quality-of-service (QoS) requirements.

Following this initial setup, the \textit{WAB-gNB} is initialized and registers with the \textit{5GC-serving-UE}, enabling the establishment of \gls{pdu} sessions for connected \glspl{ue}. Each session is supported by a Data Radio Bearer (DRB) between the \gls{ue} and the \textit{WAB-gNB}, and by an N3 \gls{gtp} tunnel between the \textit{WAB-gNB} and the \textit{5GC-serving-UE}, which is, in turn, encapsulated within the backhaul-segment \gls{pdu} session. During setup or update procedures, the \textit{WAB-gNB} can include an identifier of the associated \textit{WAB-MT} in its messages to explicitly indicate its belonging to a WAB node. This information can be used to prevent unintended connections between two \textit{WAB-gNBs} in cascade.



When a WAB node is in motion, the \textit{WAB-MT} relies on standard \gls{ue} mobility procedures to enable seamless handovers in the backhaul segment. 
As it moves within the network, the \textit{WAB-MT} may also request the establishment or modification of backhaul-segment \gls{pdu} sessions on demand based on the \textit{WAB-gNB} traffic requirements. In addition, the \textit{WAB-gNB} may establish \rev{3GPP Xn} interface connections with neighboring gNBs to improve the handover experience.

\vspace{-0.25cm}
\section{The \gls{wab} revolution}
\label{sec:wab_approach}
\rev{
Relay-based architectures are designed to position smaller \gls{bs}-equivalent nodes closer to \glspl{ue} in order to enhance coverage.
}
Despite their conceptual appeal, such solutions have historically struggled to reach large-scale deployment. Notably, \gls{iab} and femtocell systems -- two of the most representative attempts -- have proven difficult to integrate flexibly into existing networks and have faced practical limitations in terms of deployment complexity, management, and adaptability. 
\rev{
In the following, we compare \gls{wab} with these two established technologies, highlighting the key differences that allow \gls{wab} to overcome the constraints that have limited their adoption.
}
\vspace{-0.4cm}
\subsection*{\gls{wab} vs \gls{iab}}
Conceptually, \gls{iab} shares similarities with \gls{wab}, as it provides wireless backhaul to relay nodes that, in turn, offer access to end \glspl{ue}. However, \gls{iab} nodes -- the functional counterparts of \gls{wab} nodes -- are based on the \gls{cu}/\gls{du} split and include only a \gls{du} entity for the access segment. This architecture requires full network control and tight coordination between the access and backhaul segments. In contrast, \gls{wab} nodes incorporate a full-fledged \gls{gnb} for the access segment, enabling a radically \textbf{modular design} in which the access and backhaul segment can belong to independent third-party operators. These operators may interact with varying levels of coordination, depending on the deployment scenario and specific service agreements. 

A major limitation of \gls{iab} lies in the complexity of its operational procedures, particularly those related to \gls{iab} node handovers. 
Backhaul handovers in \gls{iab} are rigid and intricate, as they must account for both rerooting within the parent node's subtree and the mobility of the parent node itself, further complicating system management~\cite{miab}.
To mitigate these issues, the \gls{3gpp} introduced the \gls{mbsr} technology in Release~18~\cite{ShapingASmarter}, extending the basic \gls{iab} architecture with an additional \gls{cu} component to enable more efficient node handovers.

By contrast, \gls{wab} was conceived with intrinsic \textbf{simplicity in mobility management}: the backhaul segment handles handovers by reusing its own legacy mobility procedures, without requiring additional architectural components or introducing complexity into the access segment. Moreover, during \gls{wab} standardization, discussions within \gls{3gpp} emphasized the avoidance of multi-hop topologies, thereby eliminating related complexities and reinforcing procedural simplicity.

In addition, while accommodating \gls{du} modules within \gls{iab} nodes requires extending the 5G protocol stack and associated functionalities, \gls{wab} nodes can be readily \textbf{implemented by interconnecting \gls{cots} devices}, such as existing \gls{gnb} and \gls{ue} implementations, through standard IP-based techniques. This approach greatly simplifies deployment and fosters cost-efficient, flexible integration of heterogeneous components.

\vspace{-0.4cm}
\subsection*{\gls{wab} vs Femtocells}
Both femtocells and \gls{wab} are included among the topological enhancements introduced in Release~19 of the \gls{3gpp} specifications~\cite{wabtechreport38799}. Although these technologies share the goal of enhancing coverage, a crucial distinction sets them apart.

Femtocells have a long history within \gls{3gpp}, with deployments primarily targeting static indoor environments where macrocell signals struggle to penetrate~\cite{femto}. Their backhaul is typically provided through conventional local broadband or user-supplied connectivity, such as satellite or fiber links. In contrast, \gls{wab} represents a new paradigm that enables seamless, \textbf{multi-technology mobile backhaul}. Through this approach, 5G access can be guaranteed in low-coverage areas, continuously and on demand, by means of \gls{wab} nodes that obtain backhaul connectivity through multiple technologies while maintaining service continuity via smooth vertical and horizontal backhaul handovers.

\vspace{-0.4cm}
\rev{
\subsection*{\gls{wab}'s shift}
The previous comparison exposes a fundamental limitation of traditional backhaul solutions: they are inherently constrained either by rigid architectures and complex coordination mechanisms, or by static designs that fail to support mobility and adaptability. \gls{wab} breaks with these paradigms by introducing a decoupled, modular architecture in which access and backhaul are independently managed, potentially by different operators. This shift unlocks \textit{unprecedented flexibility}, allowing \glspl{mrn} to embed entire networks within others, recursively and across multiple hierarchical levels. As a result, \gls{wab} enables dynamic, multi-technology backhaul selection, seamless interoperability among heterogeneous components, and streamlined deployment based on \gls{cots} solutions. Most importantly, it establishes a new operational model in which \glspl{mrn} inherently support \textit{two coordinated mobility layers}: a higher-level backhaul-node mobility and a lower-level UE mobility, both enabling seamless handovers. In this sense, \gls{wab} represents not merely an evolution of relay-based architectures, but a fundamental rethinking of how backhaul and access are composed in future mobile networks.
}
\vspace{-0.2cm}

\vspace{-0.1cm}
\section{An experimental implementation}
\label{sec:testbed}

\begin{figure*}[t]
    \centering
    \includegraphics[width=1.5\columnwidth]{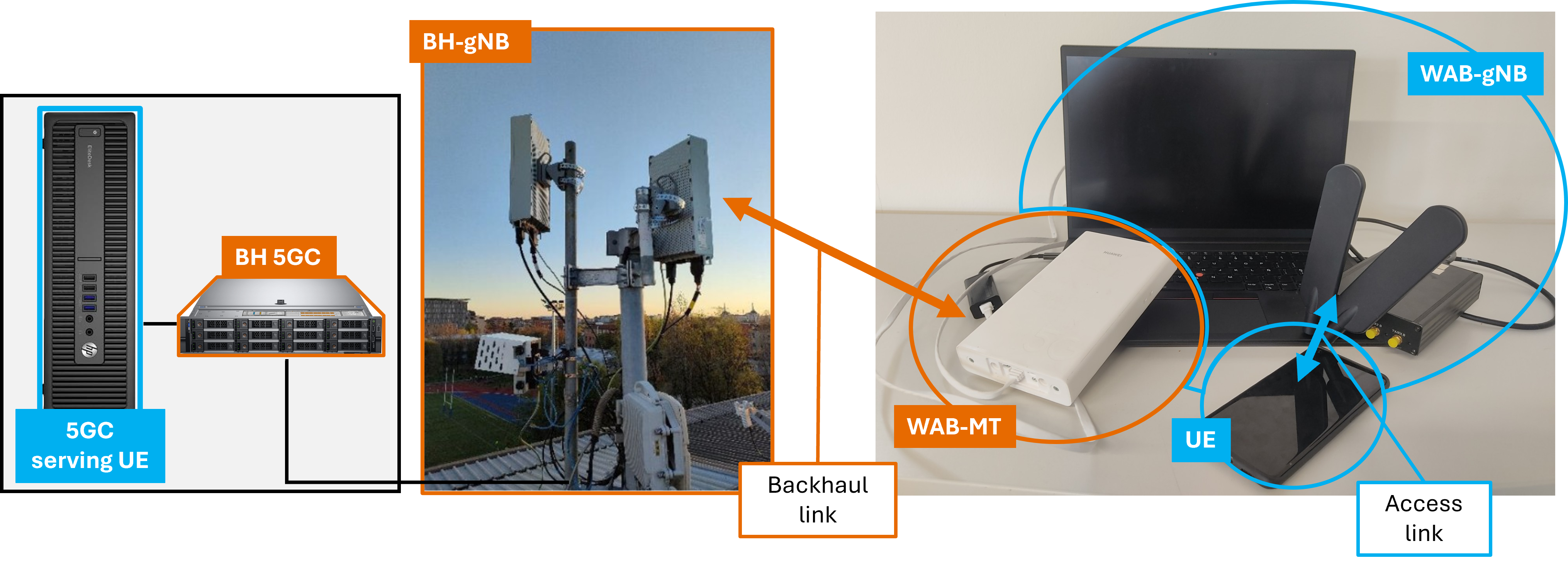}\vspace{-0.2cm}
    \caption{\rev{WAB testbed implementation: devices composing the access network are framed in light-blue, while those of the backhaul segment are framed in orange.\vspace{-0.5cm}}}
    \label{fig:testbed}
\end{figure*}

To demonstrate the feasibility of \gls{wab} systems, we present here the realization of the first practical deployment of this technology. 
The testbed follows the standard \gls{wab} architecture, with its main components illustrated in Figure~\ref{fig:testbed}.
It is implemented using a combination of commercial hardware and open-source software, ensuring full compatibility with \gls{cots} \glspl{ue}. The setup adopts an out-of-band configuration, in which the access segment operates in FR1, while the backhaul operates in FR2.

The backhaul segment leverages the FR2 5G network of the High-Frequency Campus Lab at Politecnico di Milano~\cite{hfcl}. The \textit{WAB-MT} is implemented using a \gls{cots} \gls{cpe}, connected at a carrier frequency of 27.2~GHz to the \textit{BH-gNB}, a \gls{bs} installed at a height of 22~meters on a building rooftop. The \textit{BH-gNB} is in turn connected via fiber to the \textit{BH-5GC}, a commercial \gls{5gc} instance. This FR2 network, used exclusively for research purposes, was dedicated to the testbed during all experiments and carried no external traffic. Additional system parameters are reported in~\cite{hfcl}.

The access segment provides connectivity to a \gls{cots} smartphone operating in the FR1 band. The smartphone connects to the \textit{WAB-gNB}, a \gls{sdr}-based \gls{oai} gNB~\cite{oai}, which is linked via Ethernet to the \textit{WAB-MT} providing backhaul connectivity. The \textit{5GC-serving UE} is implemented using an Open5GS core deployed on a general-purpose server connected to the \textit{BH-5GC}. \rev{Under ideal conditions, when the \gls{oai} gNB is directly connected via Ethernet to the Open5GS core, speed tests between the FR1 \gls{ue} and the core reach approximately 100~Mbps in \gls{dl} and 10~Mbps in \gls{ul}.}
 
End-to-end connectivity in the system is ensured by establishing \gls{vpn} tunnels that transport the N2 and N3 interfaces between the \textit{WAB-gNB} and the \textit{5GC-serving UE}. \rev{Specifically, we employ a WireGuard tunnel, with a modified \gls{mtu} dimension.
The use of the \gls{vpn} tunnel} is necessary because the \textit{WAB-MT}, implemented as a conventional \gls{cpe}, introduces a \gls{nat} mechanism that prevents direct  communication between the \textit{WAB-gNB} and the \textit{5GC-serving-UE}. As a result, connectivity relies on a three-layer tunneling structure: the backhaul \gls{pdu} tunnels encapsulate the \gls{vpn} tunnel, which transports the N2 and N3 access interfaces, and these in turn encapsulate the \gls{pdu} sessions of the end \gls{ue}.

This implementation demonstrates that \gls{wab} enables rapid and cost-effective deployments following standard-compliant architectures built from commercial hardware and open-source software, while maintaining full compatibility with \gls{cots} \glspl{ue}. Its flexibility and hardware independence position \gls{wab} as a practical and interoperable solution for a wide range of deployment scenarios.

 \begin{figure}[]
         \centering
      \begin{subfigure}[]{0.47\textwidth}
          \centering
           \includegraphics[width=\textwidth]{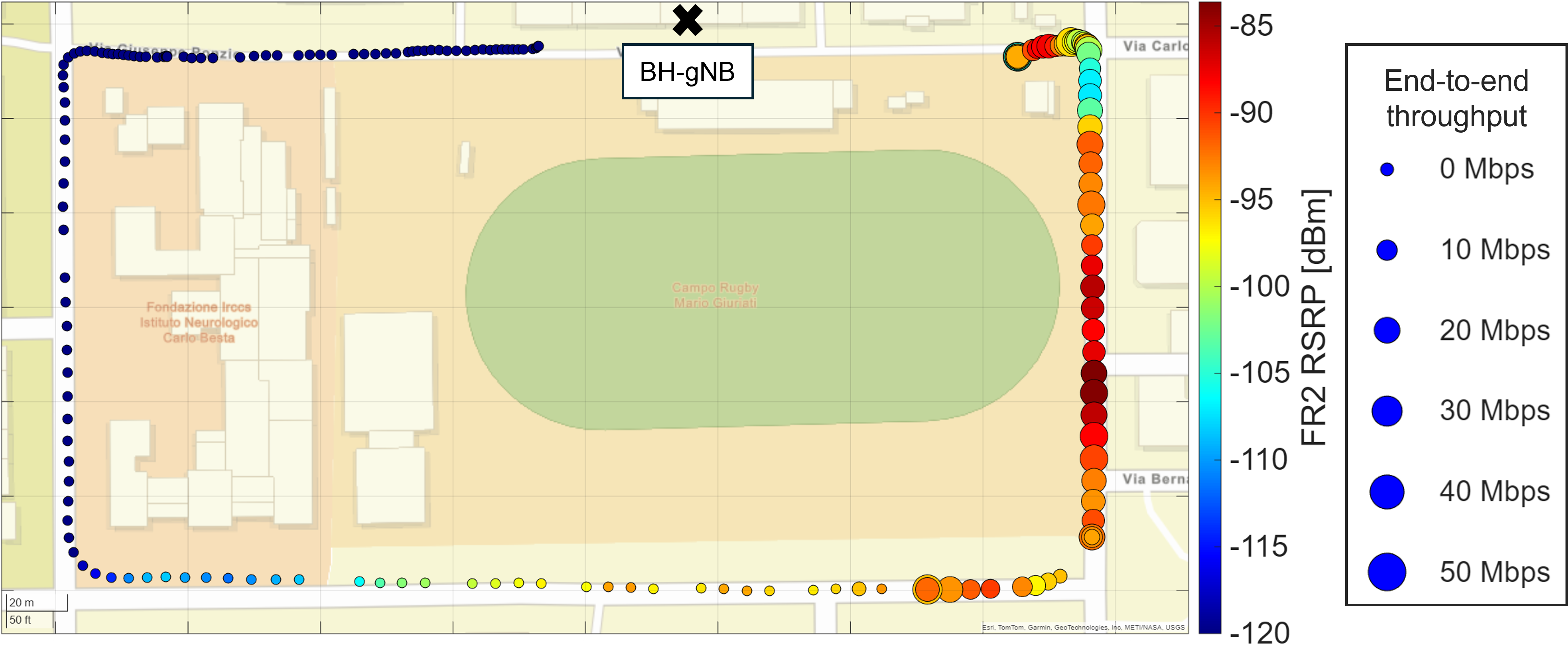}
          \centering
          \caption{\gls{dl} mobile experiments: point colors based on FR2 RSRP, point sizes based on end-to-end throughput.}
          \label{fig:mobile}
      \end{subfigure}
      \hfill
      \vspace{0.1cm}
      \begin{subfigure}[]{0.45\textwidth}
          \centering          \includegraphics[width=\textwidth]{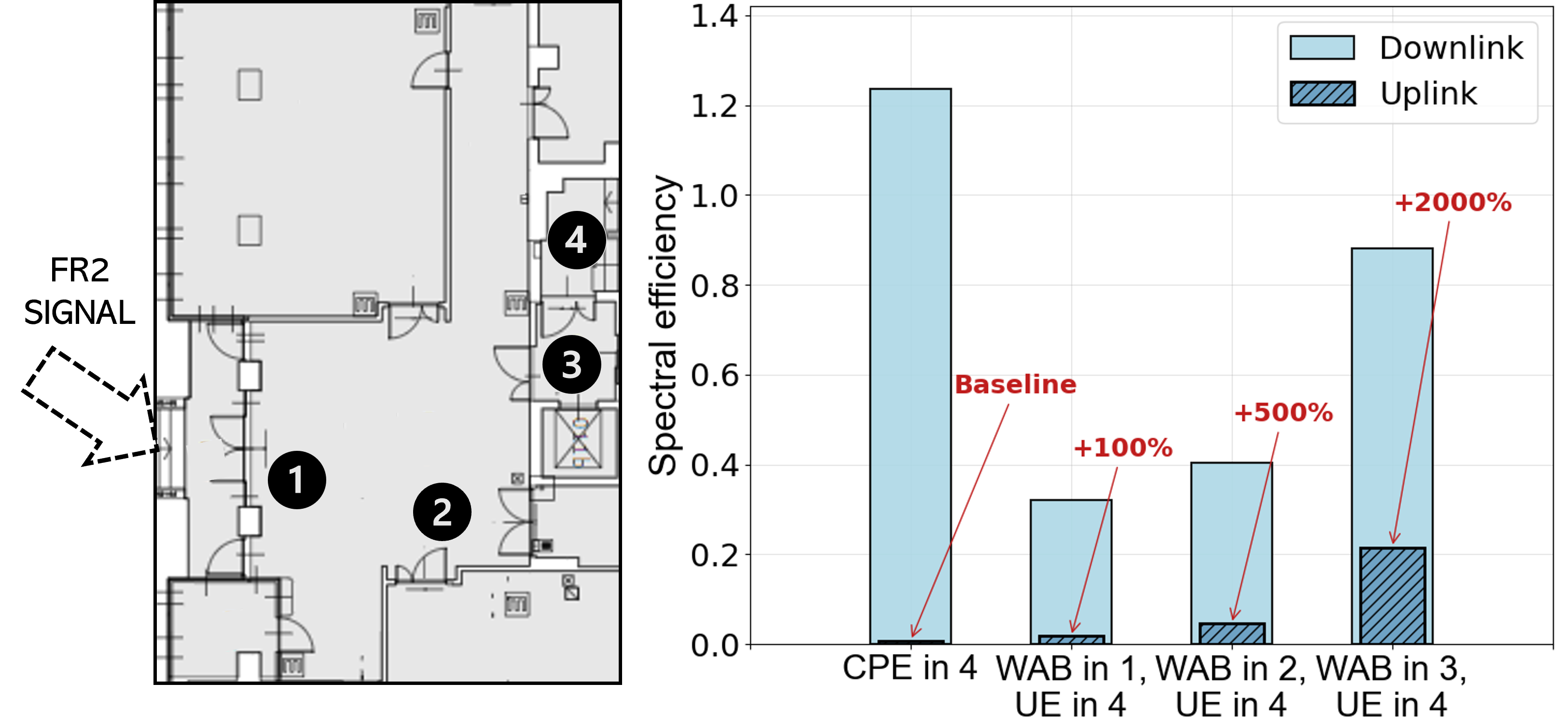}
          \caption{O2I experiments: positions of the experiments on the left, corresponding measured spectral efficiencies on the right.\vspace{-0.1cm}}
          \label{fig:o2i}
      \end{subfigure}
         \caption{Mobile and O2I \gls{wab} experiments.\vspace{-0.5cm}}
        \label{fig:results}
    
 \end{figure}

We then conducted an experimental campaign to validate the system in mobile and \gls{o2i} scenarios~\cite{rubaltelli2026enabling}. \rev{The experiments consisted of speed tests performed using \emph{iperf3} over \gls{tcp} connections.} 
In the mobile case, the \gls{wab} node was mounted on a vehicle, with the \textit{WAB-MT} installed on the rooftop and the \textit{WAB-gNB} placed inside the cabin to serve the \gls{ue}. The vehicle circulated around a sports center, traversing areas where trees and buildings intermittently obstructed the signal.  \Gls{dl} results, shown in Fig.~\ref{fig:mobile}, confirm that the end-to-end throughput closely follows the performance of the backhaul link. Under \gls{los} conditions -- characterized by the highest FR2 \gls{rsrp} -- the \gls{ue} throughput reached approximately 50~Mbps (with a 40~MHz FR1 bandwidth), and degraded progressively as the vehicle moved into \gls{nlos} regions. These results validate the correct operation of the \gls{wab} implementation, ensuring reliable service both in \gls{los} and in moderately shadowed \gls{nlos} conditions.

\rev{Additional indoor tests were conducted by placing the \gls{wab} node behind a ground-level glass façade in \gls{los} with the \textit{BH-gNB}.} Speed tests were performed with the \gls{ue} at multiple indoor locations (Fig.~\ref{fig:o2i}), and repeated by placing the FR2 \gls{cpe} at the same points to assess FR2-only performance.
\rev{Results show that FR2-only achieves higher DL spectral efficiency than the end-to-end \gls{wab} system, while in UL, especially in deep indoor conditions, \gls{wab} outperforms FR2-only. This advantage becomes more pronounced when the \gls{wab} node is moved further indoors, closer to the \gls{ue}, partially compensating for the low transmit power of the \gls{sdr}-based WAB-gNB. The histogram in Fig.~\ref{fig:o2i}, focusing on position 4, explicitly shows the increase in UL spectral efficiency compared to the FR2-only CPE case.} These results highlight \gls{wab} as a practical solution to mitigate key FR2 limitations, particularly for uplink and indoor coverage. A more extensive discussion and additional results are provided in~\cite{rubaltelli2026enabling}.

\vspace{-0.25cm}
\section{Application scenarios and Challenges}
\label{sec:application_challenges}
\begin{figure*}[h]
    \centering
    \includegraphics[width=1.4\columnwidth]{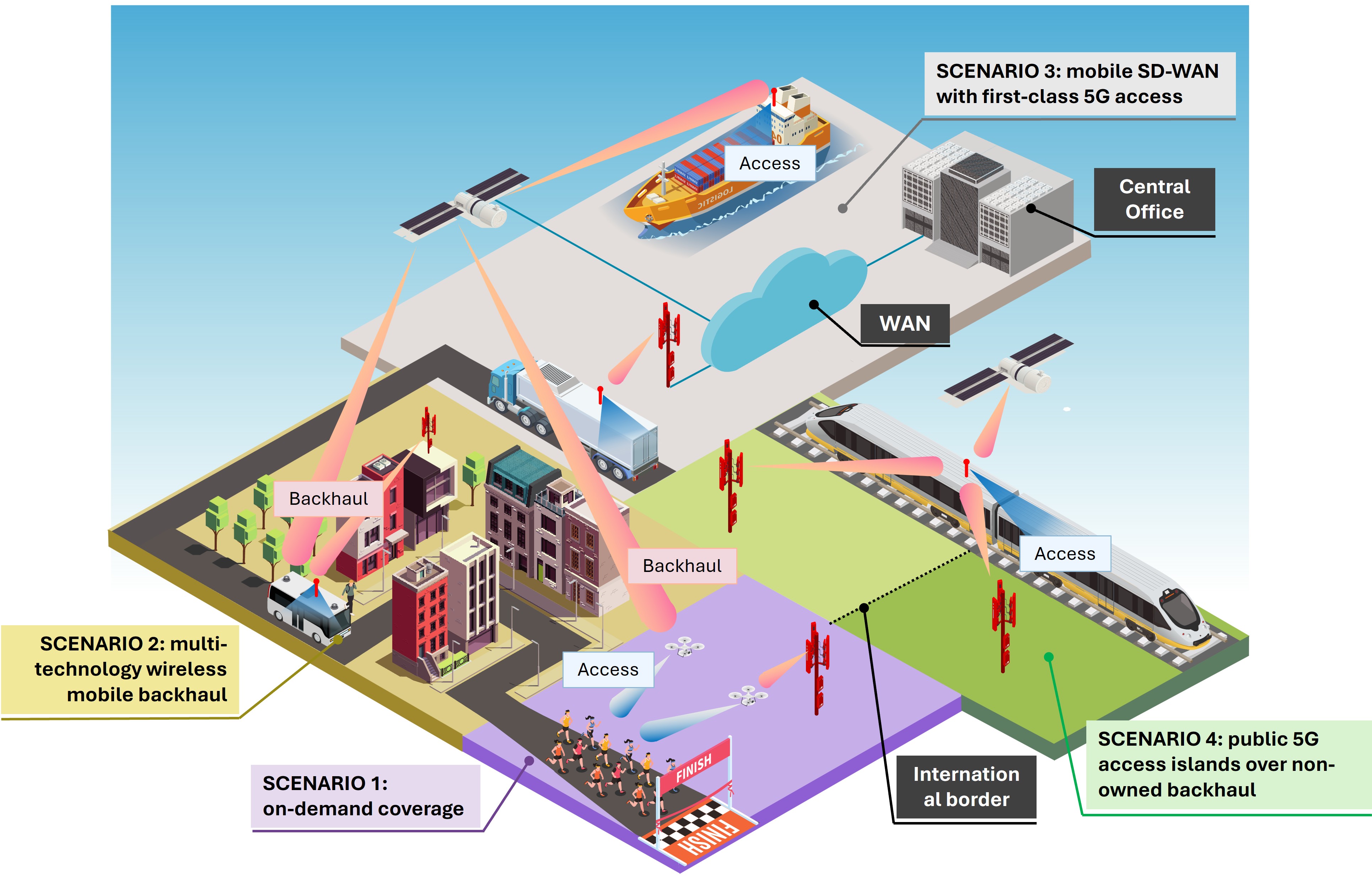}
    \caption{\rev{Application scenarios enabled by \gls{wab} technology: 1) on-demand coverage, 2) multi-technology backhaul, 3) mobile SD-WAN, 4) public 5G access islands.}\vspace{-0.6cm}}
    \label{application_scenarios}
\end{figure*}

In the evolution of mobile radio networks, several visionary application scenarios have been revisited numerous times, yet their practical realization has often proved more challenging than initially anticipated.
We argue that \gls{wab} can serve as a key enabler for four such recurrent themes, supporting intelligent and cost-efficient deployment strategies that may finally render these long-sought applications feasible in practice.

Following the presentation of these scenarios in Figure~\ref{application_scenarios}, we explore the technical challenges they introduce for \gls{wab} technologies. This discussion aims to guide future research directions and highlight how \gls{wab} could mature into a central building block of next-generation network architectures.
\vspace{-0.5cm}
\subsection*{Application Scenarios}
\subsubsection{Rapid and On-Demand Coverage}
Crowded events and emergency situations often require temporary, rapidly deployable mobile coverage in locations where fixed \glspl{bs} are impractical or unavailable. During large gatherings, such as demonstrations, marathons, or festivals, traffic surges can easily overload existing infrastructure. Likewise, natural disasters or rescue operations frequently take place in areas with limited coverage and without access to traditional backhaul.

\Gls{wab} enables fast and flexible deployment of mobile access nodes. While trailer-mounted \glspl{bs} exist, they typically require dedicated backhaul, and involve complex setup procedures. In contrast, mobile \gls{wab} nodes can be mounted on vehicles, aircraft, or vessels, dynamically obtaining backhaul from multiple technologies, including terrestrial \glspl{bs}, non-\gls{3gpp} APs, or satellites, and form reconfigurable topologies as needed. This capability allows operators and public-safety agencies to rapidly restore, or augment, coverage for nomadic users and people in distress, without relying on fixed infrastructure.

\subsubsection{Wireless Mobile Backhaul through Multiple Technologies}
Wireless backhaul has long been available, with various technologies proposed to meet evolving connectivity needs. However, solutions based on mmWave links and next-generation satellite systems often rely on custom-designed and costly solutions, which hinders large-scale deployment 
in areas where terrestrial coverage is unavailable, such as onboard aircraft, or cruise ships.

In this context, \gls{wab} can serve as a nomadic gateway for commercial \glspl{ue}, providing 5G access while multiple backhaul technologies operate transparently in the background. This includes horizontal backhaul handovers, across different nodes while in motion, and vertical handovers, across distinct backhaul technologies. Through this, \gls{wab} nodes can form architectures that ensure continuous and reliable access for \glspl{ue} via multi-technology mobile backhaul. This approach also benefits backhaul operators, who can collaborate with access operators to deliver services, and share revenue opportunities.

\subsubsection{Mobile \acrshort{sdwan} with First-Class 5G Access}
Traditional \glspl{sdwan} interconnect static enterprise branches through wired, or fixed wireless links, and therefore intrinsically lack mobility support. With \gls{wab}, organizations in logistics, public transportation, or fleet management can extend \glspl{sdwan} to nomadic branches, seamlessly connecting them to corporate networks while providing private 5G services to remote workers in the branch.

Beyond mobility, \gls{wab} enables modular composition of the \gls{sdwan} underlay, integrating multiple wireless backhaul technologies while enabling 5G in the access segment. Thanks to the flexibility of the 5G radio interface, a single \gls{wab} node can offer first-class radio access for diverse needs in the remote mobile branch -- ranging from IoT services, to industrial low-latency applications, and broadband connectivity -- using a unified access platform. By leveraging satellite communications, and backhaul mobility, it becomes possible to deploy 5G networks onboard trains, buses, airplanes, or cargo ships, integrating them into the enterprise \gls{sdwan} and enabling access to centralized services and applications.

\subsubsection{Public 5G Access Islands over Non-Owned Backhaul Networks}
Economic viability has long been the main hurdle faced by mobile network operators in providing ubiquitous connectivity in remote areas with low population density. Interestingly, similar challenges also arise in dense urban environments, where increasing traffic demand requires additional sites, and acquiring new locations, or installing new fiber links, may be equally impractical.

From a perspective complementary to the previous scenario, the \gls{wab} architecture enables operators to deploy public 5G access islands by leveraging non-owned backhaul networks. \Gls{wab} nodes can exploit existing wireless coverage from other mobile networks, or private infrastructures, such as those operated by municipalities, utilities, or public service providers.

This flexibility unlocks new deployment opportunities. Private networks used for fleet monitoring or public transportation could backhaul onboard \gls{wab} nodes, providing 5G access to passengers or enhancing surrounding coverage. Similarly, mobile \gls{wab} nodes mounted on vehicles or vessels traveling abroad could create a home-network 5G ``access island'' for transported customers, independent of local operator agreements.

\vspace{-0.2cm}
\subsection*{Technical Challenges}
\setcounter{subsubsection}{0}
\rev{
The application scenarios outlined above highlight the potential of \gls{wab} technology. However, realizing this potential requires addressing several technical challenges. In the following, we identify the most critical ones, along with the associated research questions and possible solution directions.
}
\vspace{-0.3cm}
\subsubsection{Radio Access Reconfiguration in Unreliable Backhaul}

\rev{
The mobile, flexible, and multi-technology backhaul that characterizes the \gls{wab} paradigm provides clear advantages, but its capacity may fluctuate due to congestion or propagation impairments. This can lead to outages at the \textit{WAB-gNB} when the available backhaul capacity is insufficient to sustain the bitrate required by the current radio access configuration (e.g., bandwidth, number of antenna ports, or numerology)~\cite{morini2025adaptive}.

A key research question is how to dynamically adapt the radio access configuration to match time-varying backhaul conditions. A promising direction is the introduction of adaptive \emph{gear-shifting} mechanisms, where access capacity is continuously tuned to the available backhaul bitrate. However, such reconfiguration incurs non-negligible delays, during which \glspl{ue} may experience service degradation or temporary disconnection. This raises further questions on when to trigger reconfiguration and which parameters to adjust to minimize service disruption.

In addition, backhaul interfaces in current 5G systems are typically designed under the assumption of near error-free transport, an assumption that does not hold in \gls{wab} deployments. This opens a research direction on how to redesign backhaul protocols to tolerate transmission errors and intermittent connectivity. Potential solutions include predictive link monitoring, multi-backhaul connectivity, and fast backhaul reselection mechanisms, which can jointly enhance robustness under highly dynamic conditions.
}

\subsubsection{Interference Management and Spectrum Sharing}

\rev{
\gls{wab} nodes are expected to be rapidly deployed and dynamically positioned within existing wireless environments, introducing coexistence challenges that are not typically encountered in traditional 5G \glspl{mrn}. When the \textit{WAB-gNB} and surrounding \glspl{gnb} belong to the same operator, interference may arise due to imperfect time and frequency synchronization. A non-ideal backhaul can impair synchronization mechanisms such as \gls{ptp} or \gls{ntp}, leading to misalignment between neighboring cells.


{\color{black} This issue is particularly critical in \gls{tdd} systems, where even small UL/DL configuration deviations may cause severe interference.  The effect is further exacerbated by Doppler shifts under high-speed \gls{wab} mobility. Emerging applications such as \gls{isac} are even more sensitive to desynchronization~\cite{ISAC}, potentially experiencing greater degradation than conventional interference.}

A central research question is how to enable \gls{wab} nodes to autonomously detect and align with existing synchronization patterns, while ensuring seamless coexistence with legacy infrastructure and other \gls{wab} nodes.

Interference challenges also emerge from spectrum usage. In in-band deployments, access and backhaul share the same resources, leading to self-interference. In out-of-band or multi-RAT scenarios, backhaul links may still suffer from external interference generated by heterogeneous systems. This motivates research on flexible and adaptive spectrum-sharing mechanisms tailored to \gls{wab}.

Promising solution directions include both centralized and distributed coordination approaches, possibly leveraging frameworks such as \gls{cbrs} or \gls{rim}, as well as the design of lightweight, ad-hoc interference mitigation techniques suitable for highly dynamic and mobile \gls{wab} deployments.
}

\subsubsection{Handover Optimization}

\rev{
A distinctive feature of \gls{wab} systems is that mobile \gls{wab} nodes may move together with the group of \glspl{ue} they serve. In such cases, handovers are avoided, as \glspl{ue} remain attached to the same NR Cell Identity (NCI). Conversely, when static \glspl{ue} are served by mobile \gls{wab} nodes, such as UAVs, public transport vehicles, or fleet assets, the serving node may change frequently, leading to large-scale and highly dynamic handover events.

This duality introduces a fundamental research question: how to design handover mechanisms that remain efficient under both synchronized mobility (node and UE moving together) and asynchronous mobility (mobile nodes serving static users).

Potential solutions include mobility-aware handover strategies that adapt to the movement patterns of \gls{wab} nodes, reduce signaling overhead, and minimize service interruption. Tight integration between handover management and interference coordination is also a key design guideline to ensure stable coexistence with legacy infrastructure.
}

\subsubsection{Smart Network Management and Planning}

\rev{
The dynamic and heterogeneous nature of \gls{wab} deployments opens new research directions in autonomous and intelligent network management. In particular, embedding \gls{ai} capabilities within \gls{wab} nodes can enable real-time self-organization, adaptive control, and improved robustness in complex environments. However, these solutions must operate under strict hardware and energy constraints, as \gls{wab} nodes are expected to be low-cost and resource-limited.

A key research question is how to design lightweight, distributed \gls{ai} models that can operate efficiently at the network edge while maintaining high performance.
A promising solution direction is the integration of \gls{wab} with advanced \gls{ran} architectures such as Open-\gls{ran} or AI-RAN. This would enable interoperability, flexible deployment, and the use of standardized control loops for tasks such as backhaul selection, interference mitigation, and access reconfiguration.

Finally, the inherent mobility of \gls{wab} nodes challenges conventional radio planning methodologies. Static planning approaches are no longer sufficient in networks combining fixed infrastructure with mobile, on-demand capacity providers.
This motivates research on dynamic and adaptive planning frameworks capable of continuously optimizing network configuration in response to node mobility, evolving topologies, and time-varying traffic demands.
}

\vspace{-0.3cm}
\section{Conclusion}
\label{sec:conclusion}
\rev{This article introduced the \gls{wab} architecture as a key enabler for next-generation relay-based networks, highlighting its role not simply as an evolution of existing solutions, but as a fundamental shift toward a modular and decoupled access–backhaul paradigm. This redesign enables unprecedented flexibility, including multi-technology backhaul and seamless mobility at both backhaul and access levels.

A practical implementation based on commercial hardware and open-source software demonstrated the feasibility of \gls{wab} deployments, while four representative application scenarios illustrated its potential to unlock use cases that have long remained impractical with conventional approaches.

At the same time, \gls{wab} introduces new system-level challenges that define a clear research agenda, including adaptive radio access under unreliable backhaul, mobility-aware handover strategies, scalable interference management, and lightweight AI-based mechanisms for network management and control.

As Release~19 standardization progresses, \gls{wab} stands as a pivotal enabler for 5G-Advanced and a promising foundation for future 6G network architectures.}

\vspace{-0.3cm}

\bibliographystyle{IEEEtran}
\bibliography{bibliography.bib}

%

\vspace{-1.2cm}
\begin{IEEEbiographynophoto}{Chiara Rubaltelli} is currently a PhD student at the Department of Electronics, Information and Bioengineering at Politecnico di Milano, Milan, Italy. She received her B.Sc. degree in Electronic Engineering from Università di Modena e Reggio Emilia in 2022
and the M.Sc. degree in Telecommunication Engineering from Politecnico di Milano in 2024. Her research interests include mobile radio networks, with a focus on high-frequency radio access networks, and relay-based networks.
\end{IEEEbiographynophoto}
\vspace{-1.2cm}
\begin{IEEEbiographynophoto}{Marcello Morini} is currently a PhD student at Politecnico di Milano, Department of Electronics, Information and Bio engineering. His research interests are wireless networks, with a focus on high-frequency radio access networks and smart propagation environments. He received his BSc degree in Electronic Engineering in 2020 at Università di Modena e Reggio Emilia and he completed his MSc in Telecommunication Engineering in 2022 at Politecnico di Milano. 
\end{IEEEbiographynophoto}
\vspace{-1.2cm}
\begin{IEEEbiographynophoto}{Eugenio Moro} is currently a Senior System Engineer at Qualcomm. He was an Assistant Professor at Politecnico di Milano, Milan, Italy from 2023 to 2025. He received the M.Sc. and Ph.D. cum laude in 2019 and 2023, respectively. 
His research interests are wireless networks, with a focus on high-frequency radio access, wireless network programmability and optimization and smart propagation environments.
\end{IEEEbiographynophoto}
\vspace{-1.2cm}
\begin{IEEEbiographynophoto}
{Ilario Filippini} is an Associate Professor at Politecnico di Milano.
He works on networking topics, his main research activities include radio resource management and optimization in wireless networks, programmable networks, and smart radio environments. He has co-authored more than 80 peer-reviewed conference and journal papers. He is an Editor of the ELSEVIER COMPUTER NETWORKS journal and served as an Editor of the IEEE TRANS. ON MOBILE COMPUTING (2020-23).
\end{IEEEbiographynophoto}
\vspace{-1.2cm}
\begin{IEEEbiographynophoto}{Antonio Capone} is a Professor at Politecnico di Milano. His main research activities include radio resource management in wireless networks, traffic management in software defined networks, network planning, and optimization. He is author of over 250 publications on these topics. He is an Editor of the IEEE TRANS. ON MOBILE COMPUTING, ELSEVIER COMPUTER NETWORKS, and ELSEVIER COMPUTER COMMUNICATIONS and served as an Editor for the ACM/IEEE TRANS. ON NETWORKING from 2010 to 2014.
\end{IEEEbiographynophoto}


\end{document}